# High-power intracavity single-cycle THz pulse generation using thin lithium niobate

Yicheng Wang,[1,*] Tim Vogel,[1] Mohsen Khalili,[1] Samira Mansourzadeh,[1] Kore Hasse,[2] Sergiy Suntsov,[2] Detlef Kip,[2] Clara J. Saraceno[1]

[1]*Photonics and Ultrafast Laser Science (PULS), Ruhr-Universität Bochum, Universitätsstraße 150, 44801 Bochum, Germany*
[2]*Experimental Physics and Material Sciences, Helmut-Schmidt-Universität, Holstenhofweg 85, 22043 Hamburg, Germany*
*Corresponding author: yicheng.wang@ruhr-uni-bochum.de

**Ultrafast laser driven, single-cycle THz pulsed sources hold immense potential for scientific and industrial applications; however, their limited average power hinders their widespread application. In particular, applications where high repetition rates in the multi-MHz region and beyond are required are more severely affected, due to the lower pulse energies available for frequency conversion. In this respect, resonant enhancement both in passive and active resonators is a well-known technique for boosting the efficiency of nonlinear frequency conversion; however, this route has remained poorly explored for the generation of broadband THz pulses due to the inadequacy of typically employed nonlinear crystals. Here, we demonstrate that thin lithium niobate crystals used intracavity of multimode diode-pumped mode-locked thin-disk lasers are a promising platform to circumvent these difficulties. Using a 50-μm thin lithium niobate plate intracavity of a compact high-power mode-locked thin-disk laser, we generate milliwatt-level broadband THz pulses with a spectrum extending up to 3 THz at 44.8 MHz repetition rate, driven by 264 W of intracavity average power. This approach opens the door to efficient high-power single-cycle THz generation using affordable nonlinear crystals at very high repetition rates, scalable to kilowatt-level driving power with low cost and complexity.**

Terahertz time-domain spectroscopy (THz-TDS) has become the cornerstone of many breakthrough experiments in scientific research [1–3], and is also increasingly deployed in industrial application fields, for example, for non-destructive testing [4,5]. At the front-end of every THz-TDS is a phase-stable source of single-cycle THz pulses generated via down-conversion of an ultrafast laser, typically operating in the near infrared. Naturally, progress in the performance of the THz source in a time-domain spectrometer is critical in advancing these applications, and parameters like pulse energy, bandwidth and conversion efficiency have consistently improved throughout the years.

One parameter that has seen comparatively slow progress is the average power in THz-TDS which remained low for many decades– i.e. THz pulse energy was typically traded for pulse repetition rate and vice-versa. Recent progress in high-average power Yb-based lasers [6,7] has dramatically changed this situation, with lab-scale systems now approaching the watt-level average power regime [8–10]. Among the various techniques used in these recent efforts, optical rectification (OR) in $\chi^{(2)}$ crystals is by far the most commonly explored, due to simplicity and wide availability of commonly used crystals. These recent efforts have been mostly successful in the repetition rate region <1 MHz, where mJ energies are now commonly available at hundreds of watts of average power. However, many applications, for example in imaging, would benefit from even higher repetition rates of tens of MHz and beyond, where reaching high driving pulse energies requires prohibitively high average powers. This scales price and complexity of the driving lasers, or sacrifices conversion efficiency. Very generally, most attempts so far reported to increase THz average power have simply focused on increasing the average power of the driving laser, while little effort to explore paths to enhance the efficiency of the generation process itself.

Resonant enhancement is a well-known alternative to boost the conversion efficiency and thus significantly reduce complexity and cost, that has been extensively used for nonlinear optics both in cw, nanosecond, and ultrafast regimes [11–13]. For ultrashort laser pulses, this technique is particularly attractive in the high repetition rate region well above 1 MHz, where reaching sufficient pulse energy to drive inefficient nonlinear conversion processes is challenging. For example, spectacular results have been obtained in femtosecond passive enhancement cavities and inside thin-disk lasers (TDLs) to generate XUV radiation via high harmonic generation at repetition rates of several tens of MHz [14,15].

In contrast to the large success in generating XUV light, resonantly enhanced THz generation has not seen the same progress, and early works failed to surpass the advantages of extracavity methods. Attempts were made to employ enhancement cavities (ECs) for driving THz generation using a bulk lithium niobate (LN) crystal in a Cherenkov-radiation type geometry [16]. However, passive ECs are very sensitive to loss and thus the presence of the nonlinear crystal did not allow for significant driving power buildup beyond a few watts.

The presence of gain, i.e. driving the nonlinear conversion process inside of an optical parametric oscillator (OPO) or mode-locked laser, is in this respect beneficial, however the nonlinear conversion medium needs to be chosen very carefully to maintain the delicate oscillator and/or mode-locking stability. Early on, periodically-inverted gallium arsenide (GaAs) was placed in an OPO, resulting in mW-level THz average power, at the cost of multi-cycle narrowband THz emission [17] due to limited quasi-phase matching acceptance bandwidths. Photoconductive switches were placed inside a mode-locked Yb fiber oscillator at 135 mW intracavity power, achieving 4.2 µW of THz power [18]. Such a photoconductive switch requires a rather complex fabrication, and the intracavity driving laser power is limited in ultrafast fiber oscillators. Efforts were also made with Ti:sapphire oscillators, again with limited success [19–21], where a maximum of 7 µW of THz average power was achieved, using electrically biased low-temperature GaAs as an emitter [20]. More recently, an Yb-bulk oscillator was used with gallium phosphide (GaP) at 22 W intracavity power to achieve 150 µW of THz power [22]. However, our recent extracavity study [23] shows that, at high average power, nonlinear and linear absorption in GaP introduce strong thermal effects, resulting in an axicon lens for the near-infrared (NIR) pump laser, which acts as a tremendous source of loss for the oscillator. Nonlinear absorption is well known to act as an 'inverse' saturable absorber for mode locking and can also cause mode-locking instabilities and pulse break up [24,25]. Thus, significantly higher average power operation cannot be expected with GaP. In this respect, LN is a more promising material to apply higher average power in the multi-100-W regime: it does not suffer from strong multi-photon absorption (MPA) and has a significantly higher damage threshold and effective nonlinearity. However, due to the large group velocity mismatch (GVM) between the THz and the NIR pulse, the most common scheme so far used is the tilted pulse front method [26], which is complex, sensitive to alignment and generally not suitable for intracavity use. The use of thinner LN plates has been employed at very high energies at 5 Hz repetition rate, demonstrating 0.71 mJ THz energy, however with multicycle pulses [27]. Using significantly thinner LN on substrates, another emerging area of research is the use of thin LN in a waveguided geometry for integrated THz photonics [28].

Here we propose and implement the use of thin LN plates for high-power intracavity OR. For this purpose, we choose thin-disk oscillators as they are the only technology allowing for multi-kW intracavity powers, achieving orders of magnitude higher intracavity pulse energies and peak powers than any other oscillator platform at high repetition rate [29,30], making them an ideal platform for intracavity THz generation. We operate a LN plate intracavity of a Kerr-lens mode-locked (KLM) TDL at 265 W of intracavity average power and achieve 1.3 mW of THz average power at 44.8 MHz, surpassing the state-of-the-art results of intracavity THz generation by an order of magnitude. This is achieved with a multimode pump power of 314 W, effectively trading ultrashort pulse average power for multimode cw power, thus significantly reducing the cost and complexity of the setup.

The crystals used in this experiment are fabricated from a 3-inch 50 µm x-cut 5-mol. % MgO:LiNbO$_3$ wafer (congruently melting LN, from NanoLN). The crystal thickness of 50 µm is chosen to avoid strong GVM in a collinear geometry between the driving laser at 1030 nm and the generated THz radiation. Thinner crystals could possibly be used to achieve even broader THz bandwidth but by sacrificing THz conversion efficiency, as well as at the cost of more engineering efforts for fabrication and handling of the crystals. The wafer is cut into dimensions of 17 mm(z) × 15 mm(y). This allows us to approach the maximum nonlinear coefficient $d_{33}$ of the crystal for extraordinary polarization input. A crucial aspect for the efficient use of LN for intracavity operation is the anti-reflection (AR) coating for the laser radiation wavelength, which is a very delicate task for such thin crystals. In principle, uncoated samples at Brewster's angle can be used, however this results in very reduced output coupling of the THz radiation due to internal reflections of the THz light. In our experiments, our sample is double-side AR-coated at the pump wavelength of 1030 nm using a five-layer dielectric stack deposited with a dual-magnetron sputtering machine (NanoChrome IV, IntlVac). The LN samples are further mounted on a copper frame (Fig. 1b) for intracavity experiments without additional water cooling with up to hundreds of watts.

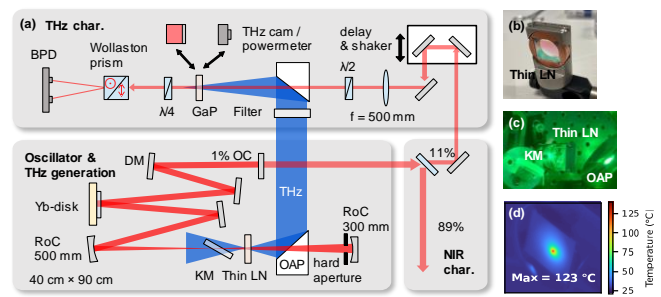

**Fig. 1.** (a) Schematic of the experimental setup. DM, dispersive mirrors. KM, Kerr medium. BPD, balanced photodetector. (b) Mounted LN sample. (c) Real setup in operation. (d) Thermal camera image of the LN.

The experimental setup is shown in Fig. 1a. The home-built KLM TDL has a separated gain material (Yb-garnet thin-disk) and Kerr medium (KM, 3-mm sapphire), offering flexibility in the design of the Kerr-lens saturable absorber and allowing for a wide range of operation parameters. The laser resonator is designed to have one focus between two concave mirrors with a radius of curvature (RoC) of 500 mm and 300 mm, respectively. Both KM and the LN are positioned at the focus. This has two benefits compared with separate waists for KM and LN: It reduces the system complexity and enhances stability when operating in ambient air at >100 W intracavity power. The stability improvement is due to the reduced self-phase modulation (SPM) turbulence from the air at the waists. The LN is placed close to the waist with a calculated beam radius of 160 µm. The position of the plate was chosen to optimize conversion efficiency following guidelines to obtain optimal conversion efficiency, which we detail in the supplemental document. A hard aperture with diameter of 3.3 mm is placed close to the end mirror. The pulse circulating intracavity fulfills the soliton law: a total round-trip group delay dispersion of -10000 fs$^2$ is introduced to compensate for the nonlinear phase within one roundtrip, originating mostly from the sapphire plate and the air inside the resonator. The thin LN plate only contributes ~3% (60 mrad) to the total SPM due to its small thickness. The cavity has a total length of 3.4 m, corresponding to a repetition rate of 44.8 MHz. One of the end mirrors of the cavity is a 1% output coupler, providing enough output power for laser diagnostics and for the electro-optic sampling (EOS) probe beam. Since we do not make use of the full 1% output power, we could close the resonator by using a higher reflectivity mirror (instead of a 1% output coupler) and reach higher efficiency (i.e. operate at lower pump power for the same intracavity power) in future experiments, offering a straightforward path for further improvements. The entire setup works in ambient air inside a dust-proof box. The oscillator and THz generation part have a compact footprint of 40 cm × 90 cm.

With a minimal pump power of 128 W at 969 nm, mode-locking is self-starting. Second harmonic generation from the LN is observed once mode-locking is initialized (Fig. 1c). As shown in Fig. 3a, mode-locking is stable with a diode pump power up to 314 W. At maximum pump power, 264 W intracavity power is achieved with a pulse duration of 115 fs, assuming a soliton pulse shape as shown in Fig. 2a. The mode-locked spectrum has a full-width at half maximum bandwidth of 10.0 nm centered at 1031.2 nm (Fig. 2b). The time-bandwidth product is 0.323, deviating only slightly from an ideal soliton pulse shape. Kelly sidebands in the optical spectrum indicate that the oscillator is operated with strongly localized dispersion and SPM in the resonator, resulting in periodic breathing in the soliton propagation [31]. As a result, the pulses at the position of the LN plate were slightly longer with 170 fs pulse duration. Radio frequency spectra with 1 GHz span and 1 MHz span are also performed indicating single-pulse operation without Q-switching instabilities (Fig. 2c,2d).

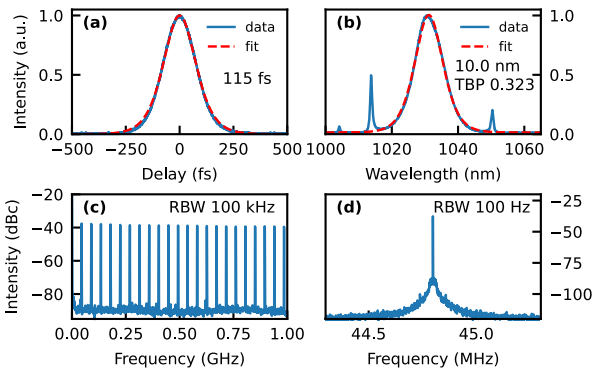

**Fig. 2.** Characterization of the KLM: (a) autocorrelation trace. (b) optical spectrum. (c) and (d) radio frequency spectra, RBW, resolution bandwidth.

The THz radiation is collected by a 50.8 mm focal length 2-inch diameter off-axis parabolic (OAP) mirror with a ~5.3 mm diameter hole along its focus axis, placed inside the cavity. A second OAP with 152.4 mm focal length is used to focus the THz light into a power meter (3A-P-THz, Ophir), a THz camera (Rigi S2, Swiss Terahertz) or onto an EOS detection crystal. To measure the THz power precisely, 4 high-density polyethylene (HDPE) plates with a total thickness of 32 mm are used to filter out the residual pump power and green second harmonic generation signal from the THz signal. As shown in Fig. 3b, we did not observe a strong saturation effect of THz power with respect to intracavity power, despite the rather high temperature on the LN sample of approximately 120 °C (Fig. 1d). The main limitation observed in our experiment is rather caused by a clamping of the intracavity power with respect to pump power as seen in Fig. 3a, likely due to thermal and other higher-order nonlinear effects in the LN plate.

At the maximum intracavity power available, we measure a maximum THz power of 510 µW at a diode pump power of 314 W and an estimated mode-locked intracavity power of 264 W. At this point, the thin LN plate operates at a peak intensity of 0.1 TW/cm$^2$ free of instabilities or damage, which is in strong contrast to the usual low intensities applied to materials such as GaP because of MPA limitations. The robustness of LN in terms of high intensity thus compensates the required small thickness to avoid GVM. At the same time operation with comparable efficiency to other commonly used materials like GaP is possible. With a measured HDPE transmission of 39%, the corrected THz power is 1.3 mW. It is important to note that the current THz power is only measured on one side of the sample. Due to the standing wave cavity design, THz is generated in both directions, which results in a doubling of the THz power. We note that in future experiments, this dual-output could be a feature of the system if considering pump-probe experiments. For other applications, the oscillator could be designed to operate in a ring-cavity configuration; this however would only be achieved at the expense of available roundtrip gain, which would also be divided by two. The THz beam is measured at the focus to have a cross section of ~1.66 × 1.37 mm$^2$, (Fig. 3b.)

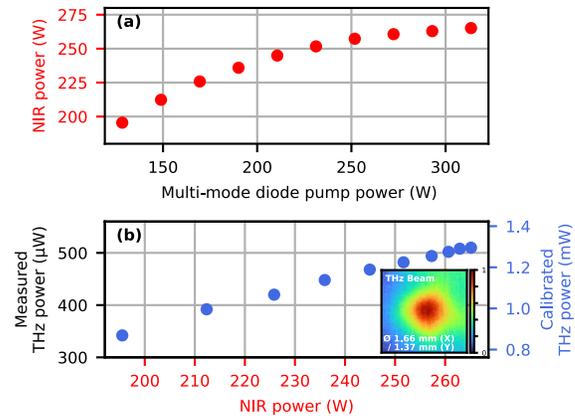

**Fig. 3.** (a) Intracavity NIR average power vs. multi-mode diode pump power. (b) THz power vs. intracavity NIR average power. Inset, THz beam profile.

Next, we measured the EOS traces at the maximum THz power. The EOS setup consists of a 1 mm GaP as detection crystal, a λ/4 plate, followed by a Wollaston prism, and a balanced photodetector. Residual pump light is filtered with a 14 mm thick HDPE plate. Dry nitrogen is used to purge the setup, reducing humidity to 20% to minimize THz absorption and improve EOS. Along the probe beamline a manual delay stage is introduced to match the temporal overlap, and a delay line with a shaker at a frequency of 10 Hz and a scanning range of 28 ps is introduced for the EOS. Within a measurement time of ~50 s, 478 traces are recorded without lock-in detection and can afterwards be averaged as shown in Fig. 4a. Notice that chopping the NIR pump beam intracavity is not possible without disturbing mode-locking. For future optimization aimed at enhancing detection via lock-in amplification, extracavity chopping of the THz beam could be considered. The noise trace is obtained by blocking the THz beam. A 2$^{nd}$-order polynomial fit is further applied to the noise trace and the THz trace to filter out the periodical offset introduced by the beam pointing of the probe, introduced by the imperfect alignment of the shaker. A 10$^{th}$-order super-Gaussian window of 23 ps (80% of the measurement data) is further applied to the time trace. The clean time trace without lock-in detection indicates a stable and high THz power.

The corresponding THz spectrum spans up to 3 THz with a dynamic range of 60 dB above the noise floor, cf. Fig. 4b. We compare the obtained traces with simulations performed by solving the coupled wave equations for OR, in a similar way to [32], including the response of the 1 mm GaP crystal, OAP/hole filtering effects [33] and echoes from the LN reflections. The low frequency part up to 1.5 THz is in a good agreement with the measurement and the dips reproduce the expected thickness of the crystal well. The small difference observed at higher frequencies is most likely due to lack of precise literature data on the THz refractive index of LN beyond ~2 THz at room temperature [34] and corresponding changes with temperature. Both the current bandwidth and output power in the setup can be optimized in the future, as high-power mode-locked TDLs with high intracavity powers in the kW range and pulses as short as 50 fs have been demonstrated [35].

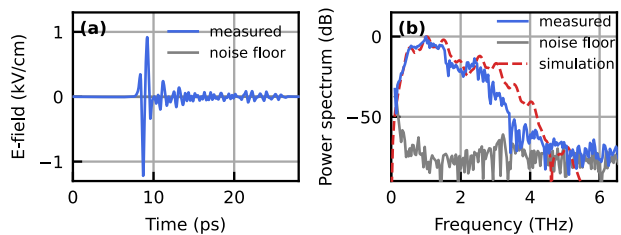

**Fig. 4.** Intracavity THz characterization. (a) THz time trace of the EOS measurement. (b) Corresponding THz spectrum with simulated spectrum.

In summary, we propose and demonstrate a new, simple and cost-effective approach for high-power THz generation using thin LN plates that is compatible with intracavity operation inside a high power TDL. In a KLM TDL operating at 264 W, 115 fs pulse duration, and at 44.8 MHz repetition rate, high THz powers of 1.3 mW (×2), with a spectrum spanning 3 THz have been obtained. We would like to highlight that these results show that thin LN is a very economic and robust platform for THz generation in general, not only for intracavity use. While ultra-thin LN films prove to be a promising platform for integrated THz applications with electric fields of V/m scale, our results using thin plates in free-space generating kV/cm fields indicate the versatility of the thin LN platform, in this form offering simple collinear, broad-bandwidth and wavelength-independent operation. The current approach has potential for further power and bandwidth scaling to tens of mW THz power in compact and efficient configuration, for example by increasing the resonator tolerance to loss with multiple passes on the disk and better thermal management of the nonlinear crystal.

**Funding.** Ruhr-Universität Bochum (Open Access Publication Funds); Ministerium für Kultur und Wissenschaft des Landes Nordrhein-Westfalen (terahertz.NRW); Deutsche Forschungsgemeinschaft (287022738 TRR 196, 390677874); HORIZON EUROPE European Research Council (805202).

**Acknowledgments.** We would like to thank Ch. Lauruschkat for his support in the production of the AR coatings.

**Disclosures.** The authors declare no conflicts of interest.

**Data availability.** Data underlying the results presented in this paper are available in Ref. [36].

**Supplemental document.** See Supplement 1 for supporting content.